# Demonstration of Optical Nonlinearity in InGaAsP/InP Passive Waveguides

Shayan Saeidi[a], Payman Rasekh[a,1], Kashif M. Awan[a,1], Alperen Tüğen[b], Mikko J. Huttunen[c], Ksenia Dolgaleva[a,d,*]

[a]*School of Electrical Engineering and Computer Science, University of Ottawa, Advanced Research Complex, 25 Templeton St., Ottawa, Ontario K1N 6N5, Canada*
[b]*Electrical and Electronics Engineering Department, Middle East Technical University, 06800, Ankara, Turkey*
[c]*Laboratory of Photonics, Tampere University of Technology, FI-33100 Tampere, Finland*
[d]*Department of Physics, University of Ottawa, Advanced Research Complex, 25 Templeton St., Ottawa, Ontario K1N 6N5, Canada*

**Abstract**

We report on the study of the third-order nonlinear optical interactions in $In_xGa_{1-x}As_yP_{1-y}$/InP strip-loaded waveguides. The material composition and waveguide structures were optimized for enhanced nonlinear optical interactions. We performed self-phase modulation, four-wave mixing and nonlinear absorption measurements at the pump wavelength 1568 nm in our waveguides. The nonlinear phase shift of up to $2.5\pi$ has been observed in self-phase modulation experiments. The measured value of the two-photon absorption coefficient $\alpha_2$ was 15 cm/GW. The four-wave mixing conversion range, representing the wavelength difference between maximally separated signal and idler spectral components, was observed to be 45 nm. Our results indicate that InGaAsP has a high potential as a material platform for nonlinear photonic devices, provided that the operation wavelength range outside the two-photon absorption window is selected.

## 1. Introduction

There has been much research effort directed towards the realization of nonlinear integrated optical devices because of their potential in all-optical signal processing [1–5]. Among different materials that have been studied for this purpose [6–10], III-V semiconductors stand out as a viable choice for passive nonlinear optical devices for two main reasons. First and most important, both active and passive integrated optical devices can potentially be combined essentially on the same material platform. This is achievable through a careful design and advanced fabrication methods, such as multilayer epitaxy and vertical tapering [11, 12]. Second, III-V semiconductors can exhibit strong optical nonlinearities accompanied by minimal nonlinear absorption achievable through a proper selection of the material composition and operation wavelength [13–15].

So far, studies of nonlinear photonic devices based on III-V semiconductors have been centered around gallium arsenide (GaAs) and related compounds such as aluminium gallium arsenide (AlGaAs) [13, 14, 16]. The range of bandgap energies associated with various material compositions of $Al_xGa_{1-x}As$ permits the use of this compound for passive nonlinear optical devices operating at the communication C-band around 1550 nm. We recently showed that there exist other interesting representatives of the group III-V that can potentially satisfy the need for nonlinear optical devices at other wavelengths [17, 18]. A variety of ternary and quaternary III-V compounds with different bandgap wavelengths can form a group of nonlinear photonic materials capable of covering the entire spectral window from ultraviolet to infrared. The present study is focused on the experimental demonstration of the nonlinear optical performance of indium gallium arsenide phosphide ($In_xGa_{1-x}As_yP_{1-y}$) integrated optical waveguides on indium phosphide (InP) substrate (we refer to them as InGaAsP/InP for conciseness).

*Corresponding author
Email address: ksenia.dolgaleva@uottawa.ca (Ksenia Dolgaleva)
[1]The authors 2 and 3 have made equal contributions to this work.

In order to maximize the occurring nonlinear optical interactions, one needs to minimize the linear and nonlinear propagation losses in waveguides. When semiconductor waveguides are transparent to the selected operation wavelength, the linear propagation loss is mostly due to scattering off from imperfections, such as epitaxial defects and waveguide surface roughness. This loss can be minimized by improving the fabrication, and can be achieved, for example, through the reduction of growth defects and plasma-induced etch roughness [19]. The dominant nonlinear loss mechanism is two-photon absorption (TPA) which can strongly contribute to the overall loss of the integrated optical device and, hence, influence the nonlinear optical performance of the device [13, 20]. In general, given a specific wavelength range of interest, it is desirable to work with the semiconductor material with the bandgap wavelength less than half of the operational wavelength. This translates into the requirement for the bandgap energy to be more than twice the photon energy, in order to minimize TPA [21, 22]. Based on this requirement, one can conclude that different III-V semiconductor compounds are suitable for nonlinear photonics at different wavelengths as they have different ranges of bandgap energies. For instance, some materials such as InGaAsP cannot have TPA suppressed at the telecom wavelengths because their bandgap energies are near the operational photon energy at 1550 nm (which corresponds to 0.8 eV). TPA coefficient of InGaAsP multi-quantum-well waveguides has been reported to be $\sim 60$ cm/GW at 1.55 $\mu$m [23], which is relatively large value. On the other hand, InGaAsP is a well established material platform for semiconductor lasers [24, 25]. This motivates one to consider the possibility of combining InGaAsP passive nonlinear optical waveguides on the same chip with InGaAsP laser sources with the potential of extending the operation ranges of the latter to longer wavelengths.

In this study, we demonstrate the potential of InGaAsP/InP waveguides for nonlinear photonic devices on-a-chip. The optimal operation wavelength range for InGaAsP passive nonlinear optical devices is at 2 $\mu$m and at longer wavelengths, as dictated by the range of its band-gap energies for various compositions that are still lattice-matched to InP substrate [18]. However, the experimental studies reported in the present work have been performed in the wavelength range between 1540 and 1590 nm. This range was chosen because some InP-based laser sources operate in this wavelength range [24, 25], and it is thus both interesting and important to study nonlinear optical performance of our InGaAsP/InP devices at these wavelengths.

Our InGaAsP/InP strip-loaded waveguides were designed specifically in a way to maximize optical nonlinearity while keeping the propagation loss relatively low [18]. On the other hand, a waveguide structure fully optimized for efficient nonlinear optical interactions would require dispersion management [26, 27] (see [18] for the corresponding designs in InGaAsP). Such dispersion-managed III-V semiconductor waveguides, frequently termed "nanowires" due to their superior compactness, require deep etching which makes fabrication of such structures a more challenging task. Moreover, the outcome is extremely sensitive to any fabrication imperfections as the sidewalls of the guiding layer in the nanowires are either exposed to the air or covered by a thin layer of a transparent dielectric as the post-processing step, either of which results in significant scattering loss. The realization of InGaAsP/InP nanowires is a separate challenge that we are currently undertaking. In the present study, we focus on the "lower-risk" structures, the shallow-etch InGaAsP/InP strip-loaded waveguides with the material composition and geometry optimized for a minimization of the effective mode area of the fundamental TE and TM modes. The goal of this design is to minimize the sensitivity of the guided modes to the fabrication imperfections while still achieving a higher efficiency of the nonlinear optical interactions, despite the presence of full material dispersion.

Here, we report on the measurement of TPA coefficient and on the observation of self-phase modulation (SPM) and four-wave mixing (FWM) in such waveguides. The most similar waveguide structures studied to date were InGaAsP multi-quantum-well rib waveguides demonstrating the nonlinear phase shift of up to $\sim 2.5\pi$ acquired through SPM at the coupled-in peak power of 3.8 W [23]. This value compares well with the nonlinear phase shift of $\pi$ reported in silicon waveguides at 60 W coupled-in peak power [28], which demonstrates the higher potential of such devices for nonlinear optical interactions compared to that of silicon waveguides. InGaAsP/InP waveguides in the present study differ from the devices reported in [23] by the geometry and by the fact that no quantum-well intermixing was performed.

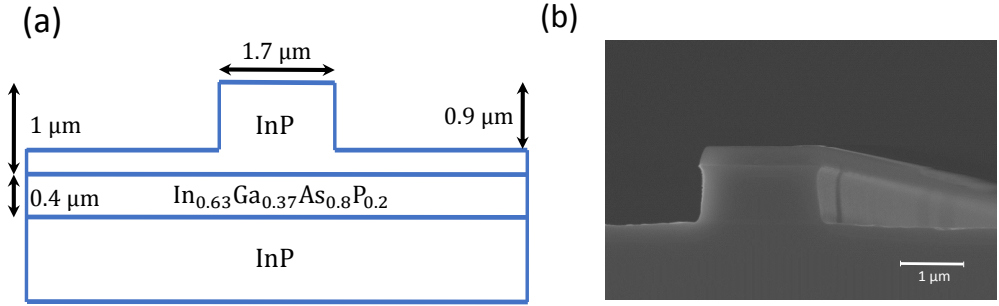

Figure 1: (a) The structure of the designed InGaAsP/InP strip-loaded waveguide. (b) SEM image of the fabricated waveguide.

## 2. Waveguide Design and Fabrication

The design of our InGaAsP/InP waveguides has been performed with the use of Lumerical Mode Solutions; the detailed description of the design process is provided in [18]. The key aspects of this design were ensuring a single-fundamental-mode operation, and the minimization of the effective mode area through a proper selection of the material composition and waveguide dimensions. In Fig. 1 (a), we show the structure of the designed waveguide. This waveguide has relatively large dimensions compared to those of more compact "nanowires" [18]. It also requires a relatively shallow etch depth, and it is known in reports as a *strip-loaded waveguide* [13]. The fundamental modes in such waveguides are well confined within the guiding layer. Therefore, the optical field does not "see" much of the fabrication imperfections, and the propagation loss is thus relatively low. The composition of the guiding layer was selected to be $In_{0.63}Ga_{0.37}As_{0.8}P_{0.2}$ with the corresponding refractive index of 3.58 at 1550 nm [29]. The refractive index of InP claddings was 3.17 [30], resulting in the index contrast of 0.41 at 1550 nm between the core and claddings. The minimal effective mode area achievable with our design was around 1.7 $\mu m^2$ at 1550 nm for a 1.7-$\mu$m-wide waveguide. The designed waveguide can potentially operate at a broad range of wavelengths spanning from the telecom C-band to around 3 $\mu$m [18].

We fabricated the waveguides using standard lithographic and etching procedures. First, the wafer was commercially grown through metalorganic chemical vapor deposition. Our target etching depth was 0.9 $\mu$m. Initial trials with electron beam resists showed that InP etching selectivity is low compared to that of the electron beam resists, hence, a hard mask was required. The wafer was first coated with 300 nm of silica using plasma-enhanced chemical vapor deposition. A 40-nm-thick layer of chromium was deposited on top of the silica layer by electron beam evaporation. The patterning of the coated wafer with waveguides was performed with a 100-kV Jeol 9500 electron beam lithography system. We used hydrogen silsesquioxane (HSQ) as the electron beam resist. Following the simulation results, we chose the waveguide width to be around 1.7 $\mu$m. Waveguide patterns were produced with the widths ranging from 1.1 to 2.1 $\mu$m, with a step size of 0.1 $\mu$m, in order to ensure that there is at least one width that corresponds to the design. In addition, operating with a range of waveguide widths can help one to experimentally verify that 1.7-$\mu$m-wide waveguide has, indeed, the best experimental performance. After patterning and developing the resist, we used the HSQ mask to transfer the pattern into the chromium layer by inductively-coupled plasma reactive-ion etching. Next, the etching of the silica layer was performed, and the waveguide pattern was transferred into the silica. Finally, the silica mask was used to transfer the waveguide pattern to InP layer. The sample was then cleaved on both sides; the resulting waveguides were 5 mm long. The scanning electron microscope (SEM) image of the fabricated waveguide cross-section is shown in Fig. 1 (b).

It must be noted here that InP etching is very challenging due to its temperature dependency. For this work, we did not have access to an etching system that supports wafer heating. To overcome this hurdle, we developed a plasma-heated etching process, where the wafer is heated by the plasma itself. This is not well-controlled and stable, hence, roughness is visible on the etched surface. This is also one of the reasons that we first pursued strip-loaded design, rather than the nanowire waveguide design, since the latter are more susceptible to roughness induced losses. For the nanowire waveguides, we are in the process of developing a plasma etching process on a system that allows wafer heating.

## 3. Optical Characterization

### 3.1. Loss Measurement

The first step in assessing the performance of an integrated optical device is the propagation loss measurement. The overall (total) loss that includes all possible loss contributions, $L_\text{t}$, can be obtained by taking a ratio between the measured optical power at the output and at the input to the waveguide. The overall loss value was measured to be around 18 dB. This value is comprised of the propagation loss $L_\text{prop}$, the coupling loss due to the mode size and shape mismatch between the free-space focused laser beam and waveguide mode $L_\text{coupl}$, and the Fresnel reflection loss $L_\text{ref}$:

$$L_\text{t} = L_\text{coupl} + 2L_\text{ref} + L_\text{prop}l. \tag{1}$$

Here $l$ represents the overall length of the waveguide. We measured the propagation loss using the Fabry-Perot loss measurement technique [31] and obtained the value of around 2.7 dB/cm for the fundamental TE mode, and 2.4 dB/cm for the fundamental TM mode for a 1.7-$\mu$m-wide waveguide. The reflectivity loss acquired by the mode at the interfaces between the semiconductor material and the air at the waveguide facets was found to be around 43%. Finally, the overall coupling loss was estimated to be $\sim$9.5 dB from Eq. (1). The values of different loss contributions agree with those observed in AlGaAs waveguides of similar geometry [13].

### 3.2. Experimental Setup

The experimental setup used for the nonlinear optical characterization of InGaAsP/InP waveguides is displayed in Fig. 2. The FWM experiments require two laser sources: a pulsed pump laser and a cw signal laser. The measurements of TPA and SPM, in turn, only require a pulsed laser. The cw laser arm is highlighted with a dashed box in Fig. 2 ("Mixing arm"): this arm was in the setup for the FWM experiments only and was not required for the rest of the nonlinear optical characterizations. We further focus on describing the setup for the FWM experiments.

We used two laser sources in our FWM experiments: a high-peak-power pulsed pump and an amplified tunable cw signal. The pump was obtained from the optical parametric oscillator (OPO), pumped by a mode-locked Ti:Sapphire laser with a repetition rate of 76 MHz. The system produced the pulsed output at 1570 nm with the pulse duration of 3 ps. The cw signal was obtained from the combination of a tunable low-power cw laser and a high-power (up to 2 W) erbium-doped fiber amplifier (EDFA), operating in the wavelength range between 1535 and 1565 nm. In each of the pump and the signal optical arms, a combination of a half-wave plate and a linear polarizer was used to independently control their polarizations. The pump and signal beams were combined at a 50% non-polarizing beam splitter, and then were coupled into the waveguide using butt-coupling with a 40$\times$ microscopic objective. We used a infrared camera (Xenics) to observe the mode profile at the output of the waveguide. The optical power was measured by the detectors with power meters, and the spectra at the input and output of the waveguide were recorded using a optical spectrum analyzer (Yokogawa AQ6370). Below we describe the results of the nonlinear optical characterization of InGaAsP waveguides.

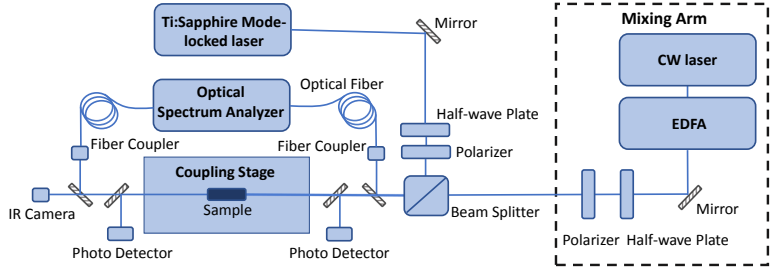

Figure 2: Experimental setup for the nonlinear optical characterization of InGaAsP/InP waveguides. FWM studies required the full setup displayed in the figure, while TPA and SPM measurements were performed without the optical arm highlighted by the dashed box (the "Mixing Arm").

3.3. Measurement of Two-Photon Absorption

In order to measure the TPA coefficient of the InGaAsP/InP waveguide, we used the nonlinear transmission method, also known as nonlinear absorption experiment [32]. Following that method, the power transmitted by the waveguide was measured as the function of the optical power at the waveguide's input. In Figure 3, we plot the transmitted power as a function of the average power launched into the waveguide for both TE (asterisks) and TM (triangles) polarizations. The characteristics at higher input powers are largely nonlinear, which indicates the presence of nonlinear absorption. Since we are operating only slightly above the bandgap (the wavelength 1568 nm falls within the region of strong TPA because the value of the bandgap wavelength corresponds to 1450 nm), this observation is intuitive.

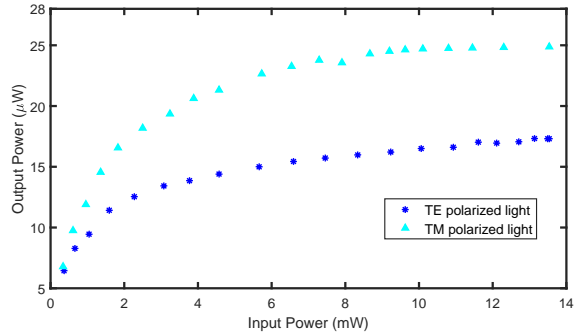

Figure 3: Output vs. input power for TE and TM polarized light. One can see the saturation behavior due to nonlinear absorption.

In order to extract the value of the TPA coefficient $\alpha_2$ from our measurements, we start with the equation for the intensity evolution in a medium with TPA [33]:

$$-\frac{dI}{dz} = \alpha I + \alpha_2 I^2. \qquad (2)$$

Here $\alpha$ represents the linear absorption coefficient, and $I$ denotes the intensity. This equation is satisfied by

$$\frac{1}{T} = e^{(\alpha L)}\left(\frac{L_{\text{eff}}}{A_{\text{eff}}}\alpha_2 P_{\text{in}} + 1\right), \qquad (3)$$

where $T$ refers to the power transmission ratio ($P_{\text{out}}/P_{\text{in}}$), $L_{\text{eff}} = [1 - \exp(-\alpha L)]/\alpha$ is the effective length,

$L$ is the length of the sample, and $A_{\text{eff}}$ is the effective mode area defined as

$$A_{\text{eff}} = \frac{\left[\int_{-\infty}^{\infty}\int_{-\infty}^{\infty}|E(x,y)|^2\,dx\,dy\right]^2}{\int_{-\infty}^{\infty}\int_{-\infty}^{\infty}|E(x,y)|^4\,dx\,dy} \quad (4)$$

in terms of the electric field $E(x,y)$ for the third-order nonlinear optical interactions [22]. The values of the effective mode area were obtained from the modal analysis performed with *Lumerical Mode Solutions* and correspond to $A_{\text{eff}} = 1.64~\mu m^2$ for the fundamental TE mode and $A_{\text{eff}} = 1.10~\mu m^2$ for the fundamental TM mode at the wavelength 1568 nm. The TPA coefficient can be obtained by measuring the power transmission ratio for different values of the input power and fitting it with Eq. (3). In Fig. 4, we plot the reciprocal transmission as a function of the coupled-in peak power (the peak power right at the entrance to the waveguide). The experimental data are shown in Fig. 4 with points, while the line represents the best line fit to these data. The slope of the line fitted to the experimental data points is proportional to $\alpha_2$. The calculated value of the TPA coefficient is $\sim 15$ cm/GW, which is approximately one quarter of the corresponding value reported for InGaAsP/InP multi-quantum-well waveguides [17]. This discrepancy can be attributed to the fact that the density of states in quantum wells near the band edges is higher compared to that of a bulk semiconductor material. Moreover, due to the difference in the material compositions in the two studies, the values of the TPA coefficients obtained in the two experiments are expected to be different.

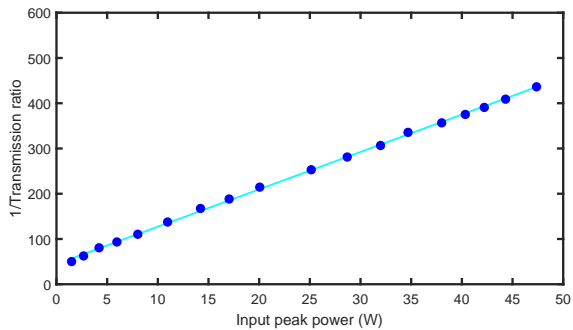

Figure 4: The TPA coefficient $\alpha_2 = 15$ cm/GW is obtained from the slope of the line fitted to the experimental data.

*3.4. Self-Phase Modulation*

As the next step in demonstrating the potential of InGaAsP nonlinear waveguides, we performed SPM experiments. For this experiment, we only used the pulsed laser source (see Fig. 2). We measured the spectra collected at the output of the waveguide at different levels of the incident power (see Fig. 5). The legend in the figure shows the values of the coupled-in peak power and the estimated nonlinear phase shift [34]. The spectral broadening with the increase of the incident power, that represents the manifestation of SPM, can be clearly observed from the graph. In our experiment, the wavelengths that we have been working with fall in the range of strong TPA, which is confirmed by the large value of the measured TPA coefficient. Free carriers, generated in the process of TPA, affect the symmetry of the spectral broadening: it becomes asymmetric due to the wavelength-selective free-carrier absorption, as confirmed by our observations (see Fig. 5). The maximum nonlinear phase shift that we were able to observe at 40 W of the coupled-in peak power was $\phi_{\text{NL}} = 2.5\pi$, as estimated from the observed spectral characteristics at the output of the waveguide [34]. Despite the strong TPA, this value is significant and indicates the presence of strong Kerr nonlinearity.

Indeed, based on our estimates of the phase shift, and using equation

$$\phi_{\text{NL}} = n_2 k_0 L I \quad (5)$$

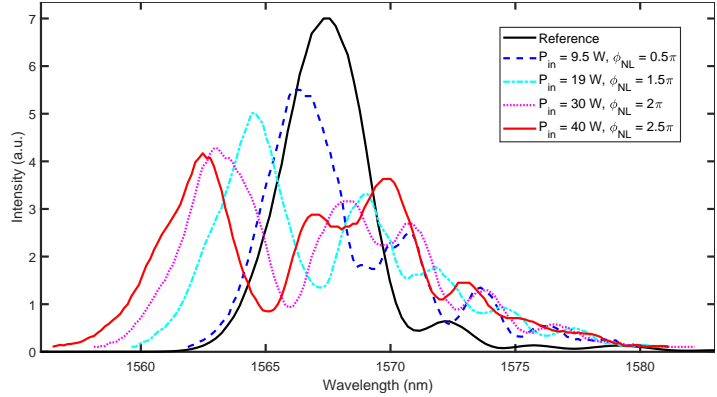

Figure 5: Spectral broadening and the nonlinear phase shift acquired by the beam propagated through the waveguide due to SPM. The legend shows the coupled-in power values and the estimated values of the nonlinear phase shift $\phi_{\text{NL}}$ of the corresponding spectra.

that relates the nonlinear phase shift $\phi_{\text{NL}}$ and the Kerr coefficient $n_2$ for SPM [20], we obtain the value $n_2 \approx 10^{-13}$ cm$^2$/W for our waveguide. We would like to emphasize the fact that this value is merely an estimate (since the nonlinear phase shift has been estimated from SPM spectra). A more precise measurement of the Kerr coefficient $n_2$ is required, which was challenging to obtain from our measurements due to the presence of ripples in the spectrum of the pulsed laser source (the pump). Nevertheless, this estimated value of $n_2$ compares well with the reported values $n_2 = 1.5 \times 10^{-13}$ cm$^2$/W for Al$_{0.18}$Ga$_{0.82}$As [22], which has been proved to be an efficient material for nonlinear photonics, and silicon $n_2 = 4.2 \times 10^{-14}$ cm$^2$/W at the telecom wavelength [35].

We now have sufficient information in order to characterize the nonlinear optical performance of the InGaAsP/InP strip-loaded waveguides in terms of the figure of merit $F$, expressed as [21]

$$F = \frac{n_2}{\lambda_0 \alpha_2}, \tag{6}$$

and the nonlinear coefficient $\gamma$, defined as

$$\gamma = \frac{2\pi n_2}{\lambda_0 A_{\text{eff}}}. \tag{7}$$

The figure of merit characterizes the potential of the material itself for nonlinear optics at the specific wavelength $\lambda_0$, while the nonlinear coefficient describes the performance of the nonlinear optical waveguide characterized by its effective mode area $A_{\text{eff}}$ at the wavelength $\lambda_0$. The nonlinear optical performance of a waveguide is optimal when both parameters are maximized, which can be achieved by a proper selection of the material (for the figure of merit), and by a proper design of the waveguide (for the nonlinear coefficient). For a nonlinear material to be considered efficient, the condition $F > 1$ must be satisfied. Based on the measured and estimated values of $n_2$ and $\alpha_2$ in our experimental study, InGaAsP has $F \approx 0.043$, which is consistent with the fact that the operation wavelengths 1530–1570 nm fall within the range of strong TPA. The nonlinear coefficient $\gamma \approx 36.85$ m$^{-1}$W$^{-1}$, characterizing the performance of the InGaAsP/InP waveguides as nonlinear devices, is almost three times larger than that of similar devices based on AlGaAs [13]. This can be attributed to the fact that we have designed our InGaAsP/InP waveguides in a way that their effective mode area is minimized to the level under 2 $\mu$m$^2$, compared to the values over 4 $\mu$m$^2$ reported in [13]. It is thus possible to exploit the full potential of such waveguides by a proper selection of the operation wavelength which, in our case, should be close to 2 $\mu$m.

### 3.5. Four-Wave Mixing

As the final step in our nonlinear optical characterizations, we have performed FWM experiments with our InGaAsP/InP waveguides. In this experiment, the whole setup, including both laser sources, was used (see Fig. 2). The pulsed pump and amplified cw beams were mixed at the beam splitter, and then were free-space-coupled into the waveguide. The values of the coupled-in power at the waveguide's entrance were estimated to be 9 W for the peak power of the pump, and 0.5 W for the average power of the cw signal. The pump laser wavelength in our experiment was centered around 1568 nm, while the cw signal, amplified with an EDFA, was tuned in the range between 1545 and 1560 nm. The spectrum at the output of the waveguide had an extra frequency component generated by the process of FWM where two pump photons with the frequency $\omega_p$ and one signal photon with the frequency $\omega_s$ interacted with the nonlinear optical medium giving rise to a single idler photon with the frequency $\omega_i$: $2\omega_p - \omega_s = \omega_i$.

In Figure 6, we show the results of the spectral measurements for different wavelength combinations of the pump and signal. The generated idler peaks corresponding to different cw signal peaks can be seen on the longer-wavelength side with respect to the pump. The results displayed in Fig. 6 were obtained for the fundamental TM modes (for both pump and signal beams): similar effect has been observed for the fundamental TE mode, but the FWM efficiency was lower in this case due to the fact that the TE mode has a larger effective mode area and, hence, exhibits a smaller nonlinear coefficient $\gamma$.

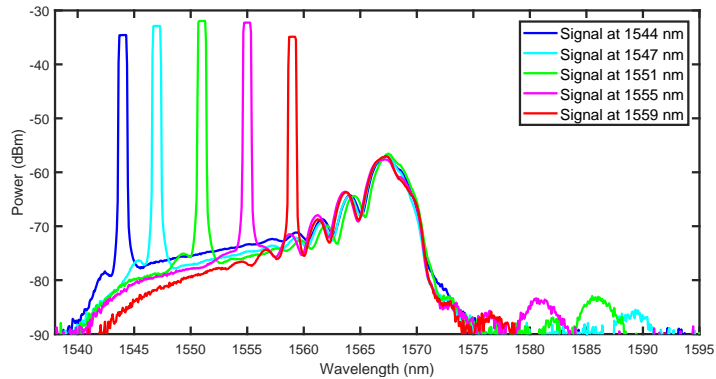

Figure 6: FWM spectra for the pump wavelength fixed at 1568 nm while the cw signal wavelengths (narrow peaks on the shorter-wavelength side) are tuned between 1545 and 1560 nm. The generated idler components corresponding to the cw peaks appear on the longer-wavelength side from the pump. The strongest idler peak ($-82.5$ dBm) is generated for the signal wavelength 1551 nm. The conversion efficiency decreases as the signal gets detuned from this wavelength.

The characteristics that we have observed in our experiment are comparable to those demonstrated with AlGaAs waveguides of similar geometry [13]. The overall conversion range (the difference in wavelength between the maximally separated signal and idler spectral components) was $\lambda_i - \lambda_s = 45$ nm, limited by the material dispersion which is the dominant dispersion mechanism in strip-loaded waveguides (see the arguments provided above and in [13]). Indeed, the material dispersion of InGaAsP in the vicinity of 1568 nm is $D_{\text{mat}} = -19500$ ps/(nm km), while the simulated overall dispersion with the waveguide contribution taken into account is around $D_{\text{tot}} = -17200$ ps/(nm km). The latter value is only slightly offset with respect to the very high value of the material dispersion and lends to the characteristic dispersion length, defined as

$$L_{\text{D}} = \frac{T_0}{D_{\text{tot}} \Delta\lambda_{\text{max}}} \quad (8)$$

in terms of the pulse duration $T_0$ and maximum wavelength separation between the pump and signal $\Delta\lambda_{\text{max}} \approx 22$ nm. The waveguides have the length $L = 5$ mm, while $L_{\text{D}} \approx 8$ mm, which is comparable with the length of the waveguide. The lack of dispersion management in our devices thus contributes to the very low efficiency and limits tuning range of the FWM process.

We next estimate the FWM conversion efficiency obtained in our experiment. Here we define the conversion efficiency as the ratio between the generated idler power and the output cw signal power, estimated from the spectral measurements. Here, the maximum conversion efficiency of around $-50$ dB was achieved. As a comparison, the value of $-38$ dB has been reported for AlGaAs nanowires [27] with engineered dispersion, ultracompact mode with $A_{\text{eff}} < 1$ $\mu$m$^2$, and managed TPA.

## 4. Conclusion

In conclusion, we report on the first measurements of the nonlinear absorption coefficient, self-phase modulation (SPM) and four-wave mixing (FWM) in InGaAsP/InP strip-loaded waveguides. We have observed the nonlinear phase shift of $2.5\pi$ in our SPM experiments, and the overall FWM conversion range of 45 nm, limited by the material dispersion and strong two-photon absorption (TPA). The TPA coefficient $\alpha_2$ was measured to be 15 cm/GW, which is already comparable to similar AlGaAs waveguide structures and could be further reduced by excitation with longer wavelengths [13]. This study represents the first set of proof-of-principle experiments demonstrating the potential of InGaAsP for nonlinear photonics on-a-chip. The next step would be the follow-up study performed at longer wavelengths such that the effect of TPA can be mitigated. This would allow one to exploit the full potential of InGaAsP for passive nonlinear optical waveguides capable of demonstrating wavelength conversion in the frequency range beyond 2 $\mu$m. It is also instructive to realize InGaAsP integrated optical waveguides with engineered dispersion in order to expand the wavelength conversion window [18]. Such waveguides, combined with integrated InGaAsP lasers, can open new spectral windows for all-optical signal processing.

To conclude, the reported study has confirmed that III-V semiconductors are well suited for nonlinear photonics. The abundance of different kinds of III-V semiconductor materials indicate that these materials can constitute a bank of excellent nonlinear optical platforms covering the entire spectral window from ultraviolet to infrared.

## Funding

This work was supported by the Canada First Research Excellence Fund award on Transformative Quantum Technologies and by the Natural Sciences and Engineering Council of Canada (Discovery Program). The additional funding sources were the Canada Research Chairs programs, and a Canadian Microelectronics Corporation award for epitaxial growth. MJH acknowledges support from the Finnish Cultural Foundation (Grant No. 00150020) and the Academy of Finland (Grant No. 308596).

## Acknowledgement

Electron-beam patterning of the InGaAsP/InP waveguides was carried out at CNF (Cornell Nanoscale Science and Technology Facility, Ithaca, NY, USA), and plasma etching was performed at QNC (Mike and Ophelia Lazaridis Quantum-Nano Centre, Waterloo, Canada).